\pdfoutput=1
\documentclass[aps,prl,superscriptaddress,twocolumn]{revtex4-2}
\usepackage[english]{babel}

\usepackage{amsmath,amsfonts}
\usepackage[normalem]{ulem}

\usepackage{bm}
\usepackage{graphics}
\usepackage{graphicx}

\usepackage[utf8x]{inputenc}
\usepackage[T1]{fontenc}
\usepackage{ae,aecompl}

\usepackage[usenames,dvipsnames]{xcolor}

\usepackage{color}
\definecolor{darkgreen}{rgb}{0,0.6,0}
\definecolor{darkblue}{rgb}{0,0,0.6}
\definecolor{darkred}{rgb}{0.6,0,0}
\definecolor{darkpurple}{rgb}{0.5,0,0.5}

\usepackage{hyperref}

\hypersetup{
bookmarksopen=true,
pdftitle=Supersymmetries in non-equilibrium Langevin dynamics,
pdfauthor=B Marguet E Agoritsas L Canet V Lecomte, 
pdftoolbar=false,               
pdfstartview={FitH},		
pdfmenubar=true,		
pdfhighlight=/O,		
colorlinks=true,		
urlcolor=darkblue,
citecolor=darkblue,		
linkcolor=darkpurple	        
}

\newcommand{\s}{\text{S}}

\newcommand{\OM}{{\textti{OM}}}
\newcommand{\MSR}{{\textti{MSR}}}
\newcommand{\SUSY}{{\textti{SUSY}}}
\newcommand{\opt}{{\text{opt}}}
\newcommand{\St}{{\textti{s}}}

\newcommand{\dd}{\!\text{d}}

\newcommand{\R}{\text{R}}
\newcommand{\ee}{\text{e}}

\newcommand{\eps}{\varepsilon}

\newcommand{\tf}{t_{\text{f}}}

\newcommand{\st}{{\text{st}}}

\renewcommand{\phi}{\varphi}

\newcommand{\p}{\partial}
\newcommand{\tr}{\operatorname{tr}}

\newcommand{\sms}{\kern.75pt}  
\newcommand{\sns}{\kern-.75pt}  

\providecommand{\textti}[1]{\text{\tiny{#1}}}

\newcommand{\hb}{\hat{h}}

\newcommand{\overbar}[1]{\mkern 1.6mu\overline{\mkern-1.6mu#1\mkern-1.6mu}\mkern 1.6mu}

\newcommand{\sectionPRL}[1]{
  \vspace*{-3.5mm}
  \section{#1}
  \vspace*{-3mm}
}

\newcommand\Psib{\overbar{\Psi}}

\newcommand{\pp}{\text{\textbf{'}}}

\providecommand{\mathds}[1]{\mathbb{#1}}

\begin{document}

\title{Supersymmetries in non-equilibrium Langevin dynamics}

\author{Bastien Marguet}
\affiliation{Institut Lumière Matière, UMR5306 Université Lyon 1-CNRS, Université de Lyon 69622 Villeurbanne, France​}
\affiliation{Université Grenoble Alpes, CNRS, LIPhy, 38000 Grenoble, France}

\author{Elisabeth Agoritsas}
\affiliation{Institute of Physics, Ecole Polytechnique Fédérale de Lausanne (EPFL), CH-1015 Lausanne, Switzerland}

\author{Léonie Canet}
\affiliation{Université Grenoble Alpes, CNRS, LPMMC, 38000 Grenoble, France}
\affiliation{Institut Universitaire de France, 1 rue Descartes, 75005 Paris, France}

\author{Vivien Lecomte}
\email[]{vivien.lecomte@univ-grenoble-alpes.fr}
\affiliation{Université Grenoble Alpes, CNRS, LIPhy, 38000 Grenoble, France}

\date{\today}

\begin{abstract}
Stochastic phenomena are often described by Langevin equations, which serve as a mesoscopic model for microscopic dynamics.
It is known since the work of Parisi and Sourlas that \emph{reversible} (or equilibrium) dynamics present supersymmetries (SUSYs).
These are revealed when the path-integral action is written as a function not only of the physical fields, but also of Grassmann fields representing a Jacobian arising from the noise distribution.
SUSYs leave the action invariant upon a transformation of the fields that mixes the physical and the Grassmann ones.
We show that, contrarily to the common belief, it is possible to extend the known reversible construction to the case of arbitrary \emph{irreversible} dynamics, 
for overdamped Langevin equations with additive white noise --~provided their steady state is known.
The construction is based on the fact that the Grassmann representation of the functional determinant is not unique, and can be chosen so as to present a generalization of the Parisi--Sourlas SUSY.
%
%
We show how such SUSYs are related to time-reversal symmetries and allow one to derive modified fluctuation-dissipation relations valid in non-equilibrium.
We give as a concrete example the results for the Kardar--Parisi--Zhang equation.
%
\end{abstract}


\maketitle

%
The dynamics of a large number of elementary constituents can often be described  by
mesoscopic stochastic equations of motion, where the effects of interactions at small scales are accounted for by friction and noise.
Such an effective Langevin~\cite{langevin1908theorie} description applies to various examples ranging from particles in a fluid to chemical or economical processes~\cite{kampen_stochastic_2007,gardiner_handbook_1994}
or cosmological inflation~\cite{starobinsky_dynamics_1982,pinol_manifestly_2020}. 

Field theory then allows one to write the probability of trajectories followed by the system
using a path-integral representation that encompasses both classical and quantum problems~\cite{zinn-justin_quantum_2002}.
The weight of a trajectory takes the form of the exponential of (\mbox{minus}) an action.
It is convenient to make the action depend not only on the physical fields, but also on non-commuting auxiliary ones --\,known as Grassmann fields\,--
representing a Jacobian arising from the noise distribution.
This action possesses a generic `supersymmetry' (SUSY), known as the BRST (Becchi--Rouet--Stora--Tyutin) symmetry~\cite{becchi_abelian_1974,tyutin_gauge_1975,becchi_renormalization_1975,becchi_renormalization_1976}.
It encodes the conservation of probability.
Also, 
when the dynamics is \emph{reversible} (\emph{i.e.}~forces derive from a potential),
a second SUSY was uncovered
by Parisi--Sourlas~\cite{parisi_supersymmetric_1982} and by Feigel'man--Tsvelik~\cite{Feigelman82}
(after a similar SUSY was found for the partition function of equilibrium problems~\cite{parisi_random_1979}).

Such SUSYs, that mix physical and Grassmann fields, look surprising in a statistical mechanical context;
yet, as other symmetries in Physics, they turned out to be a powerful tool to study a variety of problems.
These range from the dynamics of spin glasses~\cite{kurchan_replica_1991,kurchan_supersymmetry_1992},
disordered spin models~\cite{semerjian_stochastic_2004} or heteropolymers~\cite{olemskoi_supersymmetric_2001},
to finite-size effects in critical dynamics~\cite{niel_finite_1987},
localization~\cite{schwarz_supersymmetry_1997},
renormalization of the random-field Ising model~\cite{tissier_supersymmetry_2011,tarjus_random-field_2020,kaviraj_random_2020},
symmetries of Hamiltonian dynamics~\cite{gozzi_hidden_1988,gozzi_hidden_1989},
and metastability in overdamped~\cite{tanase-nicola_metastable_2004} and inertial~\cite{tailleur_kramers_2006} Langevin dynamics,
with Witten's SUSY version of Morse theory~\cite{witten_supersymmetry_1982}.
SUSYs have methodological implications for renormalization~\cite{zinn-justin_renormalization_1986}
and the derivation of variational principles~\cite{balian_static_1988}
or of the Parisi--Wu stochastic quantization~\cite{parisi1981perturbation,gozzi_functional-integral_1983,damgaard_stochastic_1987}.
%
%
The Parisi--Sourlas SUSY implies Ward identities yielding the equilibrium fluctuation-dissipation relation (FDR)~\cite{chaturvedi_ward_1984,gozzi_onsager_1984}.
When the dynamics is irreversible,
the BRST symmetry remains valid,
but the Parisi--Sourlas one is broken
\emph{e.g.}~by a driving field~\cite{zimmer_fluctuations_1993,gawedzki_critical_1986} or a colored noise~\cite{trimper_supersymmetry_1990}.
It has been argued indeed that microreversibility is at the origin of SUSY~\cite{gozzi_onsager_1984}.

In this paper, we prove the contrary,
by extending the previously known results to the case of arbitrary non-equilibrium Langevin dynamics
(in the overdamped limit and for additive Gaussian white noise).
%
We assume that the stationary distribution exists and our construction depends explicitly on it.
%
%
%
The key observation is that there are several inequivalent ways to represent the same Jacobian through Grassmann fields,
and we identify one that presents an extended SUSY generalizing the Parisi--Sourlas one.
We show that the associated Ward identities yield modified FDRs, recovering some known cases~\cite{agarwal_fluctuation-dissipation_1972,speck_restoring_2006,prost_generalized_2009}.
%
%
%
%
Then, we explain how this SUSY is directly related to a time-reversal symmetry 
between the original Langevin dynamics and its `adjoint'.
We identify the mathematical structure at the origin of the extended SUSY.
The construction can be carried out both in the
Martin--Siggia--Rose--Janssen--de~Dominicis (MSRJD) framework~\cite{Janssen1976,janssen_field-theoretic_1979,dominicis_techniques_1976,DeDominicis1978,Martin1973},
and in the  Onsager--Machlup  one~\cite{onsager_fluctuations_1953,machlup_fluctuations_1953II},
where it takes a particularly simple  form.
We finally discuss the cases of spatially correlated noise, continuum space, and 
the example of the Kardar--Parisi--Zhang (KPZ) equation~\cite{kardar_dynamic_1986}.
\smallskip

\sectionPRL{BRST SUSY}
Consider a set of scalar fields $h_i(t)$  evolving in time according to a Langevin equation
\begin{equation}
  \label{eq:LangevinDiscretefi}
  \partial_t h_i = f_i[h] + \eta_i
\end{equation}
where $f_i[h]$ is a deterministic force function of the fields $h=(h_i)$ at time $t$,
and $\eta_i(t)$ a centered Gaussian white noise 
with 
$\langle \eta_i(t)\eta_j(t') \rangle = 2 T \delta_{ij} \delta(t'-t)$
[the generalization to anisotropic correlated noise is detailed below].
For instance  $h_i(t)$  represents the spatial coordinate of a particle tagged by a discrete index $i$,
or the value of the height of an interface on a lattice site $i$ (as in the KPZ equation).
Eq.~\eqref{eq:LangevinDiscretefi} is equivalent to a Fokker--Planck evolution 
$\partial_t P[h,t]=\mathds W P[h,t]$
for the  distribution $P[h,t]$ of $h$, with
\begin{equation}
\label{eq:FokkerPlanckdiscgenericfi}
\mathds{W}\,\cdot= 
  - 
  \partial_i
  \big[
      f_i[h]
     \sns\cdot
  -
   T \partial_i \sns\cdot\sns
  \big]
\:.
\end{equation}
We denote $\partial_i\equiv\frac{\partial}{\partial h_i}$ and use  implicit summation over repeated indices (including in squares such as $X_i^2$).
We assume that the dynamics possesses a stationary distribution ${P}_{\st}[h]$ such that $\mathbb W  P_{\st}=0$,
and {define} a functional $\mathcal H[h]$ by~${P}_{\st}[h]\propto \ee^{-\frac 1T \mathcal H[h]}$.
This is the so-called quasi-potential, which exists under generic conditions~\cite{risken_fokker-planck_1996}.
Then, following Graham~\cite{graham_covariant_1977} and Eyink \emph{et al.}~\cite{eyink_hydrodynamics_1996},
we decompose the total force as the sum of a conservative force $-\partial_i \mathcal H[h]$ and a driving force $g_i[h]$ as
\begin{equation}
  \label{eq:defgifromH}
  f_i[h] 
  = 
  -\partial_i \mathcal H[h]
  + 
  g_i[h]
\,.
\end{equation}
The case of reversible dynamics is recovered for {$g_i[h]\equiv 0$}.
%
This decomposition is generic when the quasi-potential exists.
From~(\ref{eq:FokkerPlanckdiscgenericfi}), the stationary condition  $\mathbb W  P_{\st}=0$ is equivalent to an identity that will be used thoroughly:
\begin{equation}
  \label{eq:condstationdisc2}
  \partial_i g_i[h]
  = \frac 1 T 
  g_i[h]\, \partial_i \mathcal H[h]
\, .
\end{equation}
%

%
%
%
We consider the distribution of fields on a finite time window  $[0,\tf]$
and denote  $\int_t = \int_{0}^{\tf}\dd t$
(but this time window can also be~$\mathbb R$).
The path-integral representation~\cite{zinn-justin_quantum_2002} of the trajectory
probability  follows from  a mere change of variable from the Gaussian noise distribution
$\text{Prob}[\eta]\propto \ee^{-\int_t \eta_i^2 /(4T)}$
to that of the field $h$, seen from the Langevin equation~(\ref{eq:LangevinDiscretefi}) as a functional of the noise:
\begin{align}
P[h]
  =
  \Big|\frac{\delta \eta}{\delta h}\Big|
  \,
  \ee^{
   -
     \frac{1}{4T} \int_{t}\eta_i[h]^2
  }
,
\quad 
\eta_i[h]\equiv\partial_th_i-f_i[h]
\,.
\label{eq:PhOMetaih}
\end{align}
Here $\eta_i[h]$ is the expression of the noise as a function of~$h$ in the Langevin equation~(\ref{eq:LangevinDiscretefi}),
and
$ 
\big|\!\frac{\delta \eta}{\delta h}\!\big| = \big|\! \det \frac{\delta\eta_{i}[h(t)]}{\delta h_{j}\sns(t')} \big|
$
is the functional Jacobian of the change of variables from $\eta$ to $h$.
We emphasize that, 
even if the Langevin equation~(\ref{eq:LangevinDiscretefi}) is additive and does not depend on its time discretization,
 the expressions of the Jacobian and of the path-integral action do depend on the discretization chosen to write them~\cite{Langouche81,itami_universal_2017,Cugliandolo-Lecomte17a}.
We adopt the Stratonovich convention, that allows one to use the rules of calculus in the path integral~\cite{cugliandolo_building_2019},
and to reverse time without changing the discretization~\cite{arenas_supersymmetric_2012,arenas_hidden_2012}.
Following Janssen~\cite{Janssen1976},
one then linearizes the square in the exponent of~\eqref{eq:PhOMetaih} using a `response field' $\smash{\hat h_i(t)}$
to obtain the  MSRJD  action.
Introducing anticommuting Grassmann fields $\Psib_{\sns i}(t)$ and $\Psi_{\sns i}(t)$~\cite{berezin2012method} to represent 
$ 
\big|\!\frac{\delta \eta}{\delta h}\!\big|
$, we get
\begin{align}
\label{eq:PMSR}
  P[h]
  &= 
 \int 
    \mathcal D\hat h
    \mathcal D\Psi
    \mathcal D\Psib
\: \ee^{-S_\SUSY}
\\[-1mm]
\label{eq:SMSRGrass}
 S_\SUSY 
  &
  = 
 \int_t 
  \Big\{
   \hat h_i\sms\eta_i[h] - T \hat h_i^2 
   - \Psib_{\sns i} \eta_i\sns\pp[h] \Psi
  \Big\}.
\end{align}
The response field $\smash{\hat h_i}$ is integrated on the imaginary axis,
and $\eta_i\sns\pp[h] $ is the Fréchet derivative of $\eta_i[h]$
which is a linear operator acting on the vector $\Psi$ as
$
\eta_i\sns\pp[h]\Psi = \partial_j\eta_i[h]\Psi_{\sns j}
$~\footnote{
It is more rigorously defined  as $\phi_i[h+h^1]=\phi_i[h]+\phi_i[h]\pp h^1 + o(h^1)$,
implying that~$\partial_t(\phi[h])=\phi\pp\partial_t h$,
and
$\phi_i[h+\eps\Psi]=\phi_i[h]+\eps \phi_i[h]\pp \Psi$.
One has for instance for $\eta_i[h]$ defined in Eq.~(\ref{eq:PhOMetaih}):
\unexpanded{$\eta_i\sns\pp[h]\Psi=\big(\delta_{ij}\partial_t-\partial_jf_i[h]\big)\Psi_{\sns j}$}. 
}.
The BRST SUSY, which originates in the 
conservation of probability,  is a Grassmann symmetry:
it depends on a Grassmann parameter $\varepsilon$ that allows one to mix the anticommuting Grassmann and the commuting physical fields as
$h\mapsto h+\delta h$, $\hb\mapsto h+\delta \hb$, etc., 
with
\begin{equation}
\label{eq:BRS_transformation}
\text{BRST:}\
 \delta h_i = \varepsilon \Psi_{\sns i}
 \quad\:
 \delta \hb_i=0
 \quad\:
 \delta \Psib_{\sns i} = \varepsilon \hb_i
 \quad\:
 \delta \Psi_{\sns i} =0 
\,.
\end{equation}
$S_\SUSY$ is invariant under (\ref{eq:BRS_transformation}), since $\delta(\eta_i[h])=\eta_i\sns\pp[h]\delta h = \eps\sms\eta_i\sns\pp[h]\Psi$.
(We denote $\delta(X)=X[h+\delta h,...]-X[h,...]$).

\sectionPRL{Extended Parisi--Sourlas SUSY}
When forces derive from a potential ($g_i[h]\equiv 0$), another SUSY was found by Parisi--Sourlas~\cite{parisi_supersymmetric_1982} and by Feigel'man--Tsvelik~\cite{Feigelman82},
in relation with the former work of Nicolai~\cite{nicolai_supersymmetry_1980,nicolai_new_1980,nicolai_functional_1982} (see~\cite{cecotti_stochastic_1983}).
It yields the equilibrium FDR~\cite{chaturvedi_ward_1984,gozzi_onsager_1984} (as discussed below).
We now extend these results to the generic Langevin dynamics~(\ref{eq:LangevinDiscretefi}).
%
%
%
The key observation is that one can identify a Grassmann action, different from~(\ref{eq:SMSRGrass}), but that still fully represents the Langevin equation~(\ref{eq:LangevinDiscretefi})
and possesses a SUSY: 
\begin{align}
\label{eq:SSUSYdiscretedagger2}
 S_\SUSY^\dag
  &
  = 
 \!
 \int_t 
 \!
  \Big\{
   \hat h_i\sms\eta_i - T \hat h_i^2 
   +\frac 1T g_i\sms \partial_i \mathcal H
   - \Psib_{\sns i} \tilde\eta_i\sns\pp^\dag \Psi
  \Big\} 
\\
 \tilde\eta_i[h]
 &
\label{eq:defetatilde}
 =
 \partial_th_i + \partial_i\mathcal H[h] + g_i[h]
\end{align}
(for compacity we drop some dependencies in $h$).
For an operator $A$, we set $(A^\dag)_{ij}=A_{ji}$~\footnote{%
Hence, explicitly:
\unexpanded{$
\Psib_{\sns i} \tilde\eta_i\sns\pp^\dag \Psi = \Psib_{\sns i}(\delta_{ij}\partial_t +\partial_{ij}\mathcal H+\partial_i g_j)\Psi_{\sns j}
$}
}.
The Grassmann part of $S_\SUSY^\dag$ is  involving $\tilde\eta_i[h]$, 
whose signification as the noise of an `adjoint' dynamics becomes clear below when relating SUSYs to time reversal.

For a reversible dynamics ($g_i[h]\equiv 0$), one sees that  $S_\SUSY= S_\SUSY^{\smash{\dag}}$: the actions~(\ref{eq:SMSRGrass}) and~(\ref{eq:SSUSYdiscretedagger2}) are identical.
For an arbitrary irreversible dynamics ($g_i[h]\neq 0$), one has $S_\SUSY\neq S_\SUSY^{\smash{\dag}}$, and yet,
as we now show in detail, 
\emph{the actions}~(\ref{eq:SMSRGrass}) \emph{and}~(\ref{eq:SSUSYdiscretedagger2}) \emph{represent the same Langevin equation}~(\ref{eq:LangevinDiscretefi}) 
(\emph{and thus the same Fokker--Planck operator}~(\ref{eq:FokkerPlanckdiscgenericfi})).
This is due to the fact that
%
when integrating over $\Psib,\Psi$,
the extra term $\frac 1T g_i\sms \partial_i \mathcal H$ in~(\ref{eq:SSUSYdiscretedagger2}) 
ensures that the Jacobian \smash{$\big|\!\frac{\delta \eta}{\delta h}\!\big|$} is correctly represented. 
%
%
%
%
To show this,
we first recall that in Stratonovich discretization~\cite{graham_statistical_1973,LaRoTi79,janssen_field-theoretic_1979,Tirapegui82,zinn-justin_renormalization_1986,janssen_renormalized_1992,lau_state-dependent_2007,arenas_functional_2010,aron_dynamical_2016}:
\begin{equation}
  \label{eq:JacExpl}
    \Big| \dfrac{\delta \eta[h]}{\delta h} \Big|
    =
    \exp\Big\{-\frac 12 \int_t \tr f_i\pp[h]\Big\}
\end{equation}
 which can be obtained by time discretization~%
\footnote{%
Indeed,
discretizing with a time step $\Delta t$, one has 
\unexpanded{$
  \eta_i[h]_t =  \frac{h_{i,t+\Delta t}-h_{i,t}}{\Delta t} - f_i[h] \big|_{h=\frac12(h_{i,t+\Delta t}+h_{i,t})}
$}
where the time is in index (and  discretization is Stratonovich).
Hence the matrix of coordinates $(i,t;j,t')$ in the definition of the Jacobian after Eq.~(\ref{eq:PhOMetaih}) is upper triangular in the time direction (this is causality), 
so that only its equal-time components matter.
Importantly, since the time-discrete Langevin equation is read as $h_{t+\Delta t}$ function of $h_t$ and $\eta_t$,
one must pay attention that the change of variables is between $h_{t+\Delta t}$ and $\eta_t$.
Its Jacobian is thus
$\partial\eta_i[h]_t/\partial h_{j,t+\Delta t}
=
\frac{1}{\Delta t}\delta_{ij} - \frac 12 \p_j f_i[h]
$.
Factorizing by $\frac{1}{\Delta t}$ [which yields a field-independent normalization factor of the Jacobian], 
using the formula $\log \det = \tr \log$, 
one thus obtains
\unexpanded{$
  \log \big|\!\frac{\delta \eta}{\delta h}\!\big|= \sum_t \tr \log (\mathbf{1}-\frac 12 \Delta t f_i\pp[h])
$}.
Expanding at small $\Delta t$, one recovers Eq.\:(\ref{eq:JacExpl}).
},
and where the trace is $ \tr f_i\pp[h]=\partial_if_i[h]\,$.
%
%
The time discretization of $\Psib,\Psi$ in~(\ref{eq:SMSRGrass}) is crucial to correctly represent the Jacobian~(\ref{eq:JacExpl})~\footnote{%
Denoting by $X^\St_t=\frac{X_{t+\Delta t}+X_t}{2}$ the Stratonovich discretization,
\unexpanded{$
\int_t \Psib_{\sns i} \eta_i\sns\pp[h]\Psi
=
\int_t \Psib_{\sns i} (\delta_{ij}\partial_t-\p_j\sns f_i[h])\Psi_{\sns j}
$}
must be discretized as 
\unexpanded{$
\sum_t \!\Delta t\sms \Psib_{\sns i,t+\Delta t} \big(\frac{\Psi_{\sns i,t+\Delta t}-\Psi_{\sns i,t}}{\Delta t}-\p_j\sns f_i[h^\St_t]\Psi^\St_{\sns j,t}\big)
=
\sum_{tt'}\! \Psib_{\sns i,t'} M_{i,t';j,t}\Psi_{\sns j,t}
$}
with the matrix elements given by
\unexpanded{$
 M_{i,t'\!;j,t}
=
\delta_{ij} 
(\delta_{t'\!,t}-\delta_{t'\!,t+\Delta t})
-
\!\Delta t\sms
\p_j\sns f_i[h_{t'}]
\frac{\delta_{t'\!,t}+\delta_{t'\!,t+\Delta t}}{2}
$}.
As the Grassmann integral yields the determinant of $M$,
and as $M$ is triangular in the time coordinate, only the diagonal $t'=t$
matters and 
\unexpanded{$
\det M = \sum_t \det (\delta_{ij}-\!\Delta t\sms\partial_jf_i[h])
$}.
One thus recovers the Jacobian~\cite{Note3}.
}.
%
Integrating over the fields $\Psib,\Psi$ in $\ee^{-S_\SUSY^\dag}$ yields 
$|\sns\frac{\delta \tilde\eta}{\delta h}\sns|=\ee^{\frac 12\! \int_t \!\tr\sms (\p_i\mathcal H+g_i)\pp[h]}$,
see Eq.~(\ref{eq:JacExpl}).
Thus, the contribution of the last two terms in~(\ref{eq:SSUSYdiscretedagger2}) is
$
 \ee^{
 - \int_t \{ \frac 1T g_i\partial_i\sns\mathcal H - \frac 12 \tr (\partial_i\mathcal H+g_i)\pp \}
}$
but this expression is in fact equal to the Jacobian~(\ref{eq:JacExpl}),
because the stationary condition~(\ref{eq:condstationdisc2}) implies 
$\frac 1T g_i\sms\partial_i\sns\mathcal H =\tr g_i\sns\pp$.
We thus have shown
$
 \int 
    \mathcal D\hat h
    \mathcal D\Psi
    \mathcal D\Psib
\: \ee^{-S_\SUSY}
=
 \int 
    \mathcal D\hat h
    \mathcal D\Psi
    \mathcal D\Psib
\: \ee^{-S_\SUSY^\dag}
$.

%
Hence, despite being different in general, the actions $S_\SUSY$ and $S_\SUSY^{\smash\dag}$ both correctly represent the trajectory probability 
of the Langevin equation~(\ref{eq:LangevinDiscretefi})
(and we denote by $\langle\sms\cdot\sms\rangle$ and $\langle\sms\cdot\sms\rangle^{\sns\smash\dag}$ the corresponding averages).
Physically, this means that observables depending only on $h$ and $\smash{\hat h}$ have the same average:
 $\langle\mathcal O[h,\hat h ] \rangle=\langle\mathcal O[h,\hat h ] \rangle^{\sns\dag}$.
This is of course not the case if $\mathcal O\sns$ depends on $\Psi$ or $\smash{\Psib}$.
%
%
This freedom of representation originates in the fact that the Jacobian depends only on the diagonal components of the operator $\eta_i\sns\pp[h] $ 
through the trace  $\tr f_i\sns\pp[h]=\partial_if_i[h]$, and not on all its components
$(\eta_i\sns\pp[h])_j=\delta_{ij}\partial_t-\partial_jf_i[h]$~\footnote{%
This explains why one cannot transform $S_\SUSY$ into $\smash{S_\SUSY^{\smash\dag}}$:
these actions contain the same information after integrating on the Grassmann fields, but not before.
}.
%
%

Then, one checks by direct computation that
\begin{equation}
\label{eq:Ggonthefieldsdisc}
\text{PS}_1\!\sns:
\left\{
\begin{aligned}
&
  \delta h_i = \varepsilon T\sms \Psib_{\sns i}
  \qquad
  \delta \hat h_i =   \varepsilon \big(\delta_{ij}\p_t  -\p_j g_i[h]\big)\Psib_{\sns j}
\\
&
  \delta \Psi_{\sns i} = \varepsilon \big(\p_t h_i - g_i[h]- T\hat h_i\big) 
  \qquad 
  \delta \Psib_{\sns i} = 0
\end{aligned}
\right.\!\!
\end{equation}
leaves $S^{\smash{\dag}}_\SUSY$ invariant, up to time boundary terms.
This SUSY generalizes the Parisi--Sourlas one to arbitrary irreversible dynamics~(\ref{eq:LangevinDiscretefi}) since, for reversible dynamics ($g_i[h]\equiv 0$)
we have $S_\SUSY^\dag=S_\SUSY$, and~(\ref{eq:Ggonthefieldsdisc}) yields the known SUSY~\cite{parisi_supersymmetric_1982}.
An important difference with the reversible case $g_i[h]\equiv 0$ is that this transformation is now non-linear in general, because of the terms $\propto g_i[h]$ in~(\ref{eq:Ggonthefieldsdisc}).
We also uncover a dual SUSY
\begin{equation}
\label{eq:Ponthefieldsdisc}
\text{PS}_2\!:
\left\{
\begin{aligned}
 &
 \delta h_i = \varepsilon T\sms \Psib_{\sns i}
 \qquad
 \delta \hat h_i = \varepsilon   \partial_{ij} \mathcal H[h] \Psib_{\sns j}
\\
 &
 \delta \Psi_{\sns i} =- \varepsilon \big({\p_i \mathcal H[h]} - T\hat h_i\big)
 \qquad
  \delta \Psib_{\sns i} = 0
\end{aligned}
\right.
\end{equation}
which seems to have been unnoticed even for $g_i[h]\equiv 0$
(perhaps because it is non-linear, even in this case).

We emphasize that this construction can also be formulated using the superfield, 
with explicit expressions for the generators of PS$_{1,2}$~\cite{inpreparation}.
One can also transpose it to the Onsager--Machlup formalism straightforwardly:
indeed the passage from the MSRJD to the Onsager--Machlup action is done 
by integrating over the response field, which amounts
to replacing $\hat h$ by its optimal value 
$\hat h^{\opt}=\frac{1}{2T}\eta[h]$~\cite{inpreparation}.
The corresponding SUSY transformation is obtained likewise,
as made explicit below.

The non-equilibrium SUSY we derived is more intricate than in equilibrium, 
since it involves two actions ($S_\SUSY$ invariant only under BRST, and $S_\SUSY^{\smash{\dag}}$ only under PS$_{1,2}$),
and depends explicitly on the steady state.
However, it allows one to derive physical consequences, as shown now.

\sectionPRL{Modified FDRs}
Symmetries of the action imply Ward identities for correlation functions: 
denoting $h_1=h_{i_1\!}(t_1)$ (and similarly for other indices, functions or operators),
the BRST symmetry~(\ref{eq:BRS_transformation})  implies in particular
$
\langle h_1\Psib_2\rangle = 
\langle (h_1+\delta h_1)(\Psib_2+\delta\Psib_2)\rangle
$,
hence 
$
\langle h_1\delta \Psib_2\rangle +
\langle \delta h_1 \Psib_2\rangle =0
$
which means:
\begin{equation}
  \label{eq:wardPsiPsiresp}
  \langle h_1 \hat h_2 \rangle
  =
  - \langle \Psi_1 \Psib_2 \rangle
\end{equation}
and we find that the 2-point correlator of the Grassmann fields is a response function.
In particular, these correlators are 0 for $t_1<t_2$.
From the invariance of $\langle h_1\Psi_2\rangle^{\sns\dag}$ under the SUSYs PS$_{1,2}$ we similarly infer:
\begin{align}
  \label{eq:FDTmodif1}
  \big\langle
   h_1 \big(\partial_t h_2 - g_2[h_2] \big)
  \big\rangle
&  =
  T \langle h_1 \hat h_2\rangle
  -
  T \langle \Psib_1 \Psi_2 \rangle^{\sns\dag}
\\
  \label{eq:FDTmodif2}
  \big\langle
   h_1 \sms \partial_2\sns \mathcal H[h_2]
  \big\rangle
&  =
  T \langle h_1 \hat h_2\rangle
  +
  T \langle \Psib_1 \Psi_2 \rangle^{\sns\dag}
\end{align}
where we used that for observables independent of $\Psi,\Psib$, the actions $S_\SUSY$ and $S_\SUSY^{\smash{\dag}}$ yield the same averages.
The causal structure of the Grassmann contribution to $S_\SUSY^{\smash{\dag}}$ shows that 
$\langle \Psib_1 \Psi_2 \rangle^{\sns\dag}=0$ for $t_1>t_2$
\footnote {%
With the notations of~\cite{Note4}, we have that
\unexpanded{$
\langle \Psib_{i,t'} \Psi_{j,t}\rangle^{\!\dag}
=
{(M^T)^{-1}}_{i,t';j,t}
$},
%
but $M$ is lower triangular in the time direction, 
so that $(M^T)^{-1}$ is upper triangular.
This implies the causality 
\unexpanded{$
\langle \Psib_{i,t'} \Psi_{j,t}\rangle^{\!\dag} =0
$}
for $t'>t$.
}
(which can also be inferred from the interpretation of  $\langle \Psib_1 \Psi_2 \rangle^{\sns\dag}$ as a response function in the adjoint dynamics, see below).
We thus obtain two modified FDRs:
\begin{align}
  \label{eq:FDTmodif1fin}
  \big\langle
   h_1 \big(\partial_t h_2 - g_2[h_2] \big)
  \big\rangle
&  =
  T \langle h_1 \hat h_2\rangle
\qquad
\text{if }
t_1>t_2
\\
  \label{eq:FDTmodif2fin}
  \big\langle
   h_1 \sms \partial_2\sns \mathcal H[h_2]
  \big\rangle
&  =
  T \langle h_1 \hat h_2\rangle
\qquad
\text{if }
t_1>t_2
\,.
\end{align}
Note that adding~(\ref{eq:FDTmodif1}) and~(\ref{eq:FDTmodif2}), or~(\ref{eq:FDTmodif1fin}) and~(\ref{eq:FDTmodif2fin}), one obtains
%
$\langle h_1 (\eta_2[h_2]-2T\hat h_2)\rangle =0$ 
which is always valid, as can be checked using
$\delta\sns(\ee^{-S_\SUSY})/\delta \hat h_2 = (\eta_2[h_2]-2T\hat h_2)\sms\ee^{-S_\SUSY}$ and
a functional integration by part.

Since the  r.h.s.~of the relation~(\ref{eq:FDTmodif1fin}) is the response function
 $\langle h_1\hat h_2\rangle= \langle \delta h_1/\delta \mathfrak f_2 \rangle_{\mathfrak f=0}$ 
to a perturbation $f\mapsto f+\mathfrak f$ of the total force,
this relation entails a modified FDR, valid in non-equilibrium (the equilibrium one, $\langle h_1\partial_t h_2\rangle=T \langle h_1 \hat h_2\rangle$, is recovered for $g[h]\equiv 0$, and can be derived from the 
Parisi--Sourlas SUSY~\cite{chaturvedi_ward_1984,gozzi_onsager_1984}).
A relation similar  in spirit was derived in~\cite{graham_covariant_1977,eyink_hydrodynamics_1996}, but in a particular setting where the perturbation is acting only on the conservative part of the force, 
so that the l.h.s.~of~(\ref{eq:FDTmodif1fin}) has no contribution from $g[h]$.
One checks that~(\ref{eq:FDTmodif1fin})-(\ref{eq:FDTmodif2fin}) are equivalent to the Agarwal FDR~\cite{agarwal_fluctuation-dissipation_1972}
and its equivalent formulations (e.g.~\cite{falcioni_correlation_1990,risken_fokker-planck_1996,speck_restoring_2006,crooks_path-ensemble_2000,chetrite_fluctuation_2008,baiesi_fluctuations_2009,prost_generalized_2009,seifert_fluctuation-dissipation_2010,seifert_fluctuation-dissipation_2010,verley_modified_2011,dal_cengio_fluctuationdissipation_2021}).
Also, Eqs.~(\ref{eq:FDTmodif1fin})-(\ref{eq:FDTmodif2fin}) and other Ward identities can be read as providing information on the quasi-potential, when it is not known.

\sectionPRL{Structure of the extended SUSY}
Noting that $\Psib_{\sns i} (\partial_i\mathcal H+g_i)\pp[h]^{\sns\dag} \Psi = -\Psi_{\sns i} (\partial_i\mathcal H+g_i)\pp[h]\Psib $,
and that integrating by parts $\int_t \Psib_{\sns i} \partial_t \Psi_{\sns i} = \int_t \Psi_{\sns i} \partial_t \Psib_{\sns i}  $,
%
%
we define a new action $\mathbb S_\SUSY^\dag \sns =\! S_\SUSY^\dag- \frac 1 T\! \big[\mathcal H[h]\sms\big]_0^{\tf}\sns\sns $ which writes
\vspace*{-4mm}%
\begin{align}
\label{eq:SSSUSYdiscretedagger}
 \mathbb S_\SUSY^{\smash\dag}
  &
  = 
 \!\!
 \int_t 
 \!
  \Big\{
  \sns
   \hat h_i\sms\eta_i \!-\! T \hat h_i^2 
    \!-\!\frac 1T \big( \partial_t h_i \!-\! g_i\big)\partial_i \mathcal H
   \!+\! \Psi_{\sns i} \bar\eta_i\sns\pp \Psib
  \Big\} 
\\
 \bar\eta_i[h]
 &
\label{eq:defetabar}
 =
 -\partial_th_i + \partial_i\mathcal H[h] + g_i[h]
 \: .
\end{align}
%
%
With $\frac{\eta_i+\bar\eta_i}{2}=\partial_i \mathcal H$
and
$\frac{\eta_i-\bar\eta_i}{2}=\partial_t h_i - g_i$,
one obtains
%
\begin{small}%
\begin{align}%
\label{eq:SSSUSYdiscreteSYM}
\hspace*{-1.9mm}%
 \mathbb S_\SUSY^\dag
  = 
 \!\!
 \int_t 
 \!
  \Big\{
  \sns
  \!-\!
  \frac 1T
  \Big(\sns \frac{\eta_i-\bar \eta_i}{2} \!-\! T \hat h_i \!\Big)
\!
  \Big(\sns \frac{\eta_i+\bar \eta_i}{2} \!-\!  T \hat h_i\! \Big)
  \sns\!+\! \Psi_{\sns i} \bar\eta_i\sns\pp \Psib
\sns
  \Big\} \!\!\!\!\!
\end{align}%
\end{small}%
Such a rewriting renders manifest that $\mathbb S_\SUSY^{\smash{\dag}}$ is invariant under the SUSYs PS$_{1,2}\!$ (without generating any boundary term).
Indeed from~(\ref{eq:Ggonthefieldsdisc}):
%
\begin{equation}
\nonumber
  \text{PS}_1
 \Rightarrow
\left\{
 \begin{aligned}
 &
 \delta\sns\big( \tfrac{\eta_i-\bar \eta_i}{2} - T \hat h_i \big)=0,
\quad
 \delta\Psi_{\sns i} = \eps \delta\sns\big( \tfrac{\eta_i-\bar \eta_i}{2} - T \hat h_i \big)
 \\
 &
  \delta\sns\big( \tfrac{\eta_i+\bar \eta_i}{2} - T \hat h_i \big)= \eps T\bar \eta_i\sns\pp \Psib
\end{aligned}
\right.
\end{equation}
so that the variations of the two products in~(\ref{eq:SSSUSYdiscreteSYM}) cancel each other very simply.
PS$_2$ presents a similar structure with the roles of $\eta_i+\bar\eta_i$ and $\eta_i-\bar\eta_i$ exchanged.
The identified structure explains how $\partial_i\mathcal H$ and the `covariant derivative' $\partial_th-g[h]$ (see~\cite{canet2011nonperturbative} for KPZ)
play a dual role in the SUSYs PS$_{1,2}$ and in the modified FDRs~(\ref{eq:FDTmodif1})-(\ref{eq:FDTmodif2}).

The actions $S_\SUSY^{\smash{\dag}}$ and $\mathbb S_\SUSY^{\smash{\dag}}$ have an equivalent physical content as they are equal up to time-boundary terms.
A careful treatment of these shows that the averages in~(\ref{eq:FDTmodif1})-(\ref{eq:FDTmodif2}) on a finite time window are those sampled by the steady state~$P_\st[h]$ at initial time~\cite{inpreparation}.

In the Onsager--Machlup formalism, the corresponding actions are particularly simple:
$
S_\OM=\int_t\! \big\{ \frac{\eta^2}{4T}-\Psib \eta\pp \Psi  \big\}
$
and
$
\mathbb S^\dag_\OM=\int_t\! \big\{ \frac{\bar\eta^2}{4T}+\Psi \bar\eta\pp \Psib  \big\}
$,
with $S_\OM$ verifying the BRST SUSY~(\ref{eq:BRS_transformation}),
and $\mathbb S^\dag_\OM$ being invariant by the PS SUSY
$\delta h=\eps T\Psib$,
$\delta \Psi=-\frac\eps2 \bar\eta$,
$\delta \Psib=0$
corresponding to PS$_{1,2}$.

\sectionPRL{Time reversal without Grassmann}
One can also represent $P[h]$ as a path integral on the response field only,
$  P[h]
= 
 \int 
    \mathcal D\hat h
\: \ee^{-S_\MSR}
$,
with the Jacobian contribution~(\ref{eq:JacExpl}) included in the action:
\begin{align}
\label{eq:SMSRnoGrass}
 S_\MSR[h,\hat h]
  & = 
 \int_t 
  \Big\{
   \hat h_i\sms\eta_i[h] - T \hat h_i^2 
   + \tfrac 12 \partial_i f_i[h]
  \Big\}
\sms.
\end{align}
Consider a time reversal of the field
  $h_i(t) =  h^{\smash{\s}}_i\sns(t_{\R}) $
(with $t_{\R}=\tf-t$)
combined with either one of these two response-field transformations (denoting $\dot\phi=\partial_t\phi$)
\begin{align}
  \label{eq:TR1disc}
\text{TR}_1\!:\
  \hb_i(t) 
  &=
    \hb_i^{\s}\sns(t_{\R})
    -\tfrac{1}{T}
    \big(
    \dot h_i^{\s}\sns(t_{\R}) + g_i[h^\s]
    \big)
\\
\text{TR}_2\!:\
  \hb_i(t) 
  &=
  -\hb_i^{\s}\sns(t_{\R}) + \tfrac 1 T \p_i \mathcal H[h^\s]
  \: .
\end{align}
The adjoint process~\cite{crooks_path-ensemble_2000,chernyak_path-integral_2006} of~(\ref{eq:LangevinDiscretefi}) is the one with
a force $\tilde f_i[h] = -\p_i\mathcal H[h]-g_i[h]$ instead of $f_i[h]$.
It is the process followed by time-reversed trajectories~\cite{brian_anderson_reverse-time_1982}.
The action $\tilde S_\MSR$ of the adjoint process present a mapping with $S_\MSR$\sms:
\begin{equation}
  \label{eq:TRMSRgenericdisc}
  S_\MSR
  [h,\hat h] 
   =
  \tilde S_\MSR
  [h^\s,\hat h^\s] 
  -
  \tfrac 1 T \big[ \mathcal H[h^{\s}] \big]_{0}^{\tf} 
\,.
\end{equation}
To derive it, one uses the stationary condition~(\ref{eq:condstationdisc2}),
and we stress that the Jacobian and non-Jacobian contributions to the action~(\ref{eq:SMSRnoGrass}) interfere.
%
TR$_{1,2}$ imply respectively
\begin{align}
  \label{eq:FDTmodif1MSR}
  \big\langle
   h_1 \big(\partial_t h_2 - g_2[h_2] \big)
  \big\rangle
&  =
  T \langle h_1 \hat h_2\rangle
  -
  T \langle h^\R_1 \hat h^\R_2\tilde\rangle
\\
  \label{eq:FDTmodif2MSR}
  \big\langle
   h_1 \sms \partial_2\sns \mathcal H[h_2]
  \big\rangle
&  =
  T \langle h_1 \hat h_2\rangle
  +
  T \langle h^\R_1 \hat h^\R_2\tilde\rangle
\end{align}
where the superscript $\cdot\sms\vphantom{|}^{\smash \R}$ indicates that the field is time-reversed and 
$\langle\sms\cdot\sms\smash{\tilde\rangle}$ is the average for the adjoint process.
These relations imply the modified FDRs~(\ref{eq:FDTmodif1fin})-(\ref{eq:FDTmodif2fin}),
because $\langle h^\R_1 \hat h^\R_2\tilde\rangle=0$ for $t_1>t_2$ (as this response function is causal).
Note that these modified FDRs were derived above from PS$_{1,2}$, which are infinitesimal Grassmann SUSYs, in contrast to TR$_{1,2}$ which are discrete symmetries.
The mapping~(\ref{eq:TRMSRgenericdisc}) also allows one to recover that $\ee^{-\mathcal H/T}$ is the steady state~\cite{inpreparation}.
Comparing~(\ref{eq:FDTmodif1MSR})-(\ref{eq:FDTmodif2MSR}) to~(\ref{eq:FDTmodif1})-(\ref{eq:FDTmodif2}),
we also identify the Grassmann correlator $\langle \Psib_1 \Psi_2 \rangle^{\sns\dag} $ for $\mathbb S_\SUSY^\dag$ as being equal to
the time-reversed response function $\langle h^\R_1 \hat h^\R_2\tilde\rangle$ in the adjoint dynamics.
This allows one to relate such Grassmann correlators to physical correlation and response functions.

%
As we now show, this can be inferred from a BRST SUSY.
One can check by direct computation that either of the time-reversal
transformations TR$_{1,2}$ yields:
\begin{equation}
  \label{eq:SSSUSYTRtilde}
  \mathbb S^\dag[h,\hat h,\Psi,\Psib] = \tilde S_\SUSY[h^\s,\hat h^\s,-\Psib^\R,\Psi^\R]
\end{equation}
(note the exchange of $\Psi$ and $\Psib$)
where $\tilde S_\SUSY$ is the original SUSY action~(\ref{eq:SMSRGrass}) but for the adjoint process.
It possesses a BRST symmetry of the type~(\ref{eq:BRS_transformation}) from which we infer
that
$
\langle \Psib_1 \Psi_2 \rangle^{\sns\dag}
\smash{
  \stackrel{(\ref{eq:SSSUSYTRtilde})}{=}
}
-\langle \Psi_1^\R\Psib_2^\R \tilde\rangle
\stackrel{(\ref{eq:wardPsiPsiresp})}{=}
\langle h_1^\R \hat h_2^\R \tilde\rangle
$.
Hence $\langle \Psib_1 \Psi_2 \rangle^{\sns\dag}$ is
a (time-reversed) response function for the adjoint dynamics, as noted above.
Eq.~(\ref{eq:SSSUSYTRtilde}) also implies identities for higher-order correlations of $\Psi$ and $\Psib$.
%

%
%
%

\sectionPRL{Correlated noise}
For noises correlated as
$\langle \eta_i(t)\eta_j(t') \rangle = 2 T D_{ij} \delta(t'-t)$ with a symmetric invertible matrix~$D$,
 the previous results can be generalized as follows. 
Keeping the same definition for the quasi-potential $\mathcal H$,
the force is now decomposed as
$f_i=D_{ij}(-\partial_j\mathcal H+g_j)$
instead of~(\ref{eq:defgifromH}) and the stationary condition~(\ref{eq:condstationdisc2}) becomes $\frac 1T g_i D_{ij}\p_j\mathcal H=D_{ij}\p_ig_j$.
The action $S_\SUSY$ is the same with $\hat h_i^2$ replaced by $\smash{\hat h_iD_{ij}\hat h_j}$, and it verifies the BRST~(\ref{eq:BRS_transformation}).
Taking matrix notations and defining now 
$\tilde\eta=\partial_th + D(\nabla \mathcal H+g)$
and 
$\bar\eta=-\partial_th + D(\nabla \mathcal H+g)$,
the actions
\begin{align}
\nonumber
 S_\SUSY^\dag
  &
  = 
 \int_t 
  \Big\{
   \hat h \big(\eta - T D \hat h\big)
    +\frac1T gD\nabla\mathcal H
   - \Psib \sms \tilde\eta \pp^\dag\sns \Psi
  \Big\} 
\\
\nonumber
 \mathbb S_\SUSY^\dag
  &
  = 
 \!\!
 \int_t 
 \!
  \Big\{
  \sns
   \hat h\big(\eta - T D \hat h \big)
    \!-\!\frac 1T \big( \partial_t h- D g\big)\nabla \mathcal H
   + \Psi \bar\eta\pp \Psib
  \Big\} 
\end{align}
generalize~(\ref{eq:SSUSYdiscretedagger2}) and~(\ref{eq:SSSUSYdiscretedagger}),
and a factorized form similar to~(\ref{eq:SSSUSYdiscreteSYM}) can be identified~\cite{inpreparation}.
SUSYs PS$_{1,2}$ become~\footnote{%
The Onsager--Machlup actions take a simple form:
\unexpanded{$
S_\OM=\int_t\! \big\{ \frac{\eta D^{\sns-\sns 1\sns}\eta}{4T}-\Psib \eta\pp \Psi  \big\}
$}
and\,
\unexpanded{$
\mathbb S^\dag_\OM=\int_t\! \big\{ \frac{\bar\eta D^{\sns-\sns 1\sns}\bar\eta}{4T}+\Psi \bar\eta\pp \Psib  \big\}
$},
with $S_\OM$ verifying the BRST SUSY~(\ref{eq:BRS_transformation}),
and $\mathbb S^\dag_\OM$ being invariant by the PS SUSY
$\delta h=\eps T\Psib$,
$\delta \Psi=-\frac\eps2 D^{\sns-\sns 1\sns}\bar\eta$,
$\delta \Psib=0$
corresponding to PS$_{1,2}$.
}
\begin{align}
\nonumber
\!&\text{PS}_1\!\sns:
\left\{
\begin{aligned}
&
  \delta h = \varepsilon T \Psib
  \qquad
  \delta \hat h =   \varepsilon D^{-1}\sns (\p_th  - D g[h])\sns\pp\sms\Psib
\\
&
  \delta \Psi = \varepsilon D^{-1} \sns(\p_t h - D g[h]- TD\hat h) 
  \qquad 
  \delta \Psib = 0
\end{aligned}
\right.\!\!
\\[1mm]
\nonumber
&
\text{PS}_2\!:
\left\{
\begin{aligned}
 &
 \delta h = \varepsilon T  \Psib
 \qquad
 \delta \hat h = \varepsilon  (\nabla \mathcal H[h])\pp\sms\Psib
\\
 &
 \delta \Psi = - \varepsilon D^{-1}\sns(D \nabla \mathcal H[h] - T D\hat h)
 \qquad
  \delta \Psib = 0
\end{aligned}
\right.
\end{align}
and they imply the following modified FDR: 
\begin{align}
  \label{eq:FDTmodif1D}
  \big\langle
   h_1 \big(\partial_t h_2 - Dg_2[h_2] \big)
  \big\rangle
&  =
  T \langle h_1 D\hat h_2\rangle
  -
  T \langle \Psib_{\sns 1} D\Psi_2 \rangle^{\sns\dag}
\\
  \label{eq:FDTmodif2D}
  \big\langle
   h_1 \sms  \nabla\sns \mathcal H[h_2]
  \big\rangle
&  =
  T \langle h_1  \hat h_2\rangle
  +
  T \langle \Psib_{\sns 1} \Psi_2 \rangle^{\sns\dag}
.
\end{align}
One has  $\langle \Psib_1 \Psi_2 \rangle^{\sns\dag}=\langle \Psib_1 D\Psi_2 \rangle^{\sns\dag}=0$ if $t_1>t_2$.

\sectionPRL{KPZ equation and continuous space}
Choosing $\mathcal H[h]=\frac\nu 2\sum_i (\nabla_ih)^2$ [with $\nabla_ih=h_{i+1}-h_i$] and
$
g_i[h]
=
  \frac \lambda 6 
  \big[
  (\nabla_{i} h)^2
+
  \nabla_{i} h
  \nabla_{i-1} h
+
  (\nabla_{i-1} h)^2
  \big]
$,
the Langevin equation~(\ref{eq:LangevinDiscretefi}) is a discretized version of the continuum KPZ equation
$\partial_th=\nu\p_x^2h+\frac \lambda 2 (\partial_xh)^2+\eta$.
It possesses the SUSYs we have derived together with the modified FDRs,
since the chosen discretizations of $\mathcal H$ and of the non-linear term $g_i[h]$ ensure that 
both sides of the stationary condition~(\ref{eq:condstationdisc2}) is~0.
Such a situation with an orthogonal decomposition of the force ($g_i \partial_i\mathcal H=0$) 
and a zero-divergence ($\partial_i g_i=0$) could be generic~\cite{eyink_hydrodynamics_1996}.

If $i$ is a lattice index, the continuous-space limit of our results is obtained directly.
For KPZ one has for instance
$\langle h_1(\partial_t h -\frac \lambda2 (\partial_x h)^2)_2\rangle=T\langle h_1\hat h_2\rangle$
and
$\langle h_1 \partial^2_x h_2\rangle=T\langle h_1\hat h_2\rangle$ 
if $t_1>t_2$. 
The second relation was derived in~\cite{canet2011nonperturbative}.
Note that \emph{not all spatial discretizations of the non-linear term satisfy}~(\ref{eq:condstationdisc2}):
hence, in general the discretization of gradients must be specified when it comes to SUSY, FDR and time reversal, because $\partial_i g_i$ is ambiguous in the continuum if $g[h]$ depends on gradients
--~as also seen in singularities of the functional Fokker--Planck equation~\cite{inpreparation}.

%
%

\sectionPRL{Discussion and Outlook}
We have identified SUSYs related to arbitrary Langevin equations with Gaussian additive white noise,
generalizing long-known results that were restricted to reversible settings~\cite{parisi_supersymmetric_1982,Feigelman82}.
They can be expressed both in the MSRJD formalism and in  the Onsager--Machlup one.
%
%
The price to pay is an explicit dependency on the stationary state, and a more complex structure: two actions
both representing the same physical process and each presenting different SUSYs. 
 The important outcome is that they entail modified non-equilibrium FDRs~\cite{agarwal_fluctuation-dissipation_1972} (that provide information on the steady-state when it is not known).
 %
%
As illustrated for the KPZ equation, the case of spatially continuous models is obtained directly from the results we presented,
but the spatial discretization of gradients has to be specified (to make sense of $\partial_ig_i$ in the continuum).

The construction we presented is reminiscent of the derivation of the Jarzynski relation by Mallick \emph{et al.}~\cite{mallick_field-theoretic_2011}, 
and it would be interesting to find a unified framework.
Our results apply to non-equilibrium models with known steady state, such as the zero-range process~\cite{spitzer_frank_interaction_1970,levine_zero-range_2005,evans_nonequilibrium_2005}
or mass transport models~\cite{guioth_mass_2017},
and other cases~\cite{schutz_1_2001,giardina_duality_2009,frassek_non-compact_2020}.
In the small-noise limit, the adjoint dynamics is often known in Macroscopic Fluctuation Theory~\cite{bertini_macroscopic_2015}, and thus the SUSYs PS$_{1,2}$ should be applicable.
We note in general that, in the small-noise asymptotics of Langevin process~\cite{freidlin_random_2012}, the quasi-potential $\mathcal H[h]$ can become a singular (non-differentiable) function of its argument~\cite{graham_weak-noise_1984,graham_weak-noise_1985,kamenev_field_2011}, even though $\mathcal H[h]$ is regular as long as $T$ is finite.
This implies that the $T\to 0$ limit has to be taken in a careful way.
The case of non-Gaussian noise could be investigated~\cite{zinn-justin_renormalization_1986}.
The extensions to inertial Langevin equations, or singular ($D$ not invertible) or colored noise, or multiplicative noise deserve further investigations.

The SUSYs we have unveiled are defined for path integrals,
but the reversible SUSY also has an operator version, with the Fokker--Planck operator completed by fermionic operators representing the Grassmann variables;
it was used by Kurchan \emph{et al.}~to study metastability in overdamped~\cite{tanase-nicola_metastable_2004} and inertial~\cite{tailleur_kramers_2006} Langevin dynamics,
 see also~\cite{drummond_stochastic_2012}.
It would be interesting to translate our results in these settings. 
It is a non-trivial task already in the overdamped case, since in the reversible case the equality of the actions~(\ref{eq:SMSRGrass}) and~(\ref{eq:SSUSYdiscretedagger2}) 
corresponds to the fact that the extended (fermionic) Fokker--Planck operator can be made Hermitian (which is an essential aspect of Kurchan \emph{et al.}'s construction),
while the same property does not hold in the generic irreversible case that we are considering. 
Last,
it could be instructive to identify the relation between our results and the slave process of Refs.~\cite{dean_equilibrium_2005,drummond_stochastic_2012},
and more generally with cohomology~\cite{ovchinnikov_introduction_2016,wegner_supermathematics_2016}.
%
%

\bigskip
\begin{acknowledgments}
The authors thank Matthieu Tissier for very useful discussions.
E.A.~acknowledges support from the Swiss National Science Foundation by the SNSF Ambizione Grant PZ00P2{\_}173962.
L.C.~acknowledges support from the ANR-18-CE92-0019 Grant NeqFluids and support from Institut Universitaire de France.
V.L.~acknowledges financial support from the ERC Starting Grant No.~680275 MALIG, the ANR-18-CE30-0028-01 Grant LABS and the ANR-15-CE40-0020-03 Grant LSD.
\end{acknowledgments}

\bibliographystyle{plain_url}
\bibliography{KPZ_symmetries_short-new}

\begin{thebibliography}{100}

\bibitem{langevin1908theorie}
P.~Langevin.
\newblock \href {https://gallica.bnf.fr/ark:/12148/bpt6k3100t/f530} {Sur la
  th{\'e}orie du mouvement brownien}.
\newblock {\em C. R. Acad. Sci. (Paris)} {\bf 146}, 530 (1908).

\bibitem{kampen_stochastic_2007}
N.~G.~van Kampen.
\newblock {\em Stochastic processes in physics and chemistry}.
\newblock North-{Holland} personal library. Elsevier Amsterdam ; Boston 3rd
  edition 2007.

\bibitem{gardiner_handbook_1994}
C.~W. Gardiner.
\newblock {\em Handbook of stochastic methods for physics, chemistry, and the
  natural sciences}.
\newblock Number~13 in Springer series in synergetics. Springer-Verlag Berlin ;
  New York 2nd edition 1994.

\bibitem{starobinsky_dynamics_1982}
A.~A. Starobinsky.
\newblock \href {http://dx.doi.org/10.1016/0370-2693(82)90541-X} {Dynamics of
  phase transition in the new inflationary universe scenario and generation of
  perturbations}.
\newblock {\em Physics Letters B} {\bf 117}, 175 (1982).

\bibitem{pinol_manifestly_2020}
L.~Pinol, S.~Renaux-Petel, and Y.~Tada.
\newblock \href {http://arxiv.org/abs/2008.07497} {A manifestly covariant
  theory of multifield stochastic inflation in phase space}.
\newblock arXiv:2008.07497 [astro-ph.CO] 2020.

\bibitem{zinn-justin_quantum_2002}
J.~Zinn-Justin.
\newblock {\em Quantum field theory and critical phenomena}.
\newblock Number 113 in International series of monographs on physics.
  Clarendon Press ; Oxford University Press Oxford : New York 4th edition 2002.

\bibitem{becchi_abelian_1974}
C.~Becchi, A.~Rouet, and R.~Stora.
\newblock \href {http://dx.doi.org/10.1016/0370-2693(74)90058-6} {The abelian
  {Higgs} {Kibble} model, unitarity of the {S}-operator}.
\newblock {\em Physics Letters B} {\bf 52}, 344 (1974).

\bibitem{tyutin_gauge_1975}
I.~V. Tyutin.
\newblock \href {http://arxiv.org/abs/0812.0580} {Gauge invariance in field
  theory and statistical physics in operator formalism}.
\newblock Lebedev Institute preprint, arXiv:0812.0580 [hep-th] 1975.

\bibitem{becchi_renormalization_1975}
C.~Becchi, A.~Rouet, and R.~Stora.
\newblock \href {http://dx.doi.org/10.1007/BF01614158} {Renormalization of the
  abelian {Higgs}-{Kibble} model}.
\newblock {\em Commun. Math. Phys.} {\bf 42}, 127 (1975).

\bibitem{becchi_renormalization_1976}
C.~Becchi, A.~Rouet, and R.~Stora.
\newblock \href {http://dx.doi.org/10.1016/0003-4916(76)90156-1}
  {Renormalization of gauge theories}.
\newblock {\em Annals of Physics} {\bf 98}, 287 (1976).

\bibitem{parisi_supersymmetric_1982}
G.~Parisi and N.~Sourlas.
\newblock \href {http://dx.doi.org/10.1016/0550-3213(82)90538-7}
  {Supersymmetric field theories and stochastic differential equations}.
\newblock {\em Nuclear Physics B} {\bf 206}, 321 (1982).

\bibitem{Feigelman82}
M.~V. Feigel'man and A.~M. Tsvelik.
\newblock \href {http://www.jetp.ac.ru/cgi-bin/e/index/e/56/4/p823?a=list}
  {Hidden supersymmetry of stochastic dissipative dynamics}.
\newblock {\em Sov. Phys. JETP} {\bf 56}, 823 (1982).

\bibitem{parisi_random_1979}
G.~Parisi and N.~Sourlas.
\newblock \href {http://dx.doi.org/10.1103/PhysRevLett.43.744} {Random magnetic
  fields, supersymmetry, and negative dimensions}.
\newblock {\em Phys. Rev. Lett.} {\bf 43}, 744 (1979).

\bibitem{kurchan_replica_1991}
J.~Kurchan.
\newblock \href {http://dx.doi.org/10.1088/0305-4470/24/21/011} {Replica trick
  to calculate means of absolute values: applications to stochastic equations}.
\newblock {\em J. Phys. A: Math. Gen.} {\bf 24}, 4969 (1991).

\bibitem{kurchan_supersymmetry_1992}
J.~Kurchan.
\newblock \href {http://dx.doi.org/10.1051/jp1:1992214} {Supersymmetry in spin
  glass dynamics}.
\newblock {\em J. Phys. I France} {\bf 2}, 1333 (1992).

\bibitem{semerjian_stochastic_2004}
G.~Semerjian, L.~F. Cugliandolo, and A.~Montanari.
\newblock \href {http://dx.doi.org/10.1023/B:JOSS.0000019821.08230.72} {On the
  stochastic dynamics of disordered spin models}.
\newblock {\em J. Stat. Phys.} {\bf 115}, 493 (2004).

\bibitem{olemskoi_supersymmetric_2001}
A.~I. Olemskoi.
\newblock \href {http://dx.doi.org/10.1070/PU2001v044n05ABEH000921}
  {Supersymmetric field theory of a nonequilibrium stochastic system as applied
  to disordered heteropolymers}.
\newblock {\em Physics-Uspekhi} {\bf 44}, 479 (2001).

\bibitem{niel_finite_1987}
J.~C. Niel and J.~Zinn-Justin.
\newblock \href {http://dx.doi.org/10.1016/0550-3213(87)90153-2} {Finite size
  effects in critical dynamics}.
\newblock {\em Nuclear Physics B} {\bf 280}, 355 (1987).

\bibitem{schwarz_supersymmetry_1997}
A.~Schwarz and O.~Zaboronsky.
\newblock \href {http://dx.doi.org/10.1007/BF02506415} {Supersymmetry and
  localization}.
\newblock {\em Commun. Math. Phys.} {\bf 183}, 463 (1997).

\bibitem{tissier_supersymmetry_2011}
M.~Tissier and G.~Tarjus.
\newblock \href {http://dx.doi.org/10.1103/PhysRevLett.107.041601}
  {Supersymmetry and its spontaneous breaking in the {Random} {Field} {Ising}
  {Model}}.
\newblock {\em Phys. Rev. Lett.} {\bf 107}, 041601 (2011).

\bibitem{tarjus_random-field_2020}
G.~Tarjus and M.~Tissier.
\newblock \href {http://dx.doi.org/10.1140/epjb/e2020-100489-1} {Random-field
  {Ising} and {O}({N}) models: theoretical description through the functional
  renormalization group}.
\newblock {\em Eur. Phys. J. B} {\bf 93}, 50 (2020).

\bibitem{kaviraj_random_2020}
A.~Kaviraj, S.~Rychkov, and E.~Trevisani.
\newblock \href {http://arxiv.org/abs/2009.10087} {Random {Field} {Ising}
  {Model} and {Parisi}-{Sourlas} supersymmetry {II}. {Renormalization} group}.
\newblock arXiv:2009.10087 [cond-mat, physics:hep-th] 2020.

\bibitem{gozzi_hidden_1988}
E.~Gozzi.
\newblock \href {http://dx.doi.org/10.1016/0370-2693(88)90611-9} {Hidden {BRS}
  invariance in classical mechanics}.
\newblock {\em Physics Letters B} {\bf 201}, 525 (1988).

\bibitem{gozzi_hidden_1989}
E.~Gozzi, M.~Reuter, and W.~D. Thacker.
\newblock \href {http://dx.doi.org/10.1103/PhysRevD.40.3363} {Hidden {BRS}
  invariance in classical mechanics. {II}}.
\newblock {\em Phys. Rev. D} {\bf 40}, 3363 (1989).

\bibitem{tanase-nicola_metastable_2004}
S.~T{\u a}nase-Nicola and J.~Kurchan.
\newblock \href {http://dx.doi.org/10.1023/B:JOSS.0000041739.53068.6a}
  {Metastable states, transitions, basins and borders at finite temperatures}.
\newblock {\em J. Stat. Phys.} {\bf 116}, 1201 (2004).

\bibitem{tailleur_kramers_2006}
J.~Tailleur, S.~T{\u a}nase-Nicola, and J.~Kurchan.
\newblock \href {http://dx.doi.org/10.1007/s10955-005-8059-x} {Kramers equation
  and supersymmetry}.
\newblock {\em J. Stat. Phys.} {\bf 122}, 557 (2006).

\bibitem{witten_supersymmetry_1982}
E.~Witten.
\newblock \href {https://projecteuclid.org/euclid.jdg/1214437492}
  {Supersymmetry and {Morse} theory}.
\newblock {\em J. Differential Geom.} {\bf 17}, 661 (1982).

\bibitem{zinn-justin_renormalization_1986}
J.~Zinn-Justin.
\newblock \href {http://dx.doi.org/10.1016/0550-3213(86)90592-4}
  {Renormalization and stochastic quantization}.
\newblock {\em Nuclear Physics B} {\bf 275}, 135 (1986).

\bibitem{balian_static_1988}
R.~Balian and M.~V{\'e}n{\'e}roni.
\newblock \href {http://dx.doi.org/10.1016/0003-4916(88)90280-1} {Static and
  dynamic variational principles for expectation values of observables}.
\newblock {\em Annals of Physics} {\bf 187}, 29 (1988).

\bibitem{parisi1981perturbation}
G.~Parisi and Y.~S. Wu.
\newblock \href
  {https://www.openaccessrepository.it/record/18105/files/LNF_81_017.pdf}
  {Perturbation theory without gauge fixing}.
\newblock {\em Scientia Sinica} {\bf 24}, 483 (1981).

\bibitem{gozzi_functional-integral_1983}
E.~Gozzi.
\newblock \href {http://dx.doi.org/10.1103/PhysRevD.28.1922}
  {Functional-integral approach to {Parisi}-{Wu} stochastic quantization:
  scalar theory}.
\newblock {\em Phys. Rev. D} {\bf 28}, 1922 (1983).

\bibitem{damgaard_stochastic_1987}
P.~H. Damgaard and H.~H{\"u}ffel.
\newblock \href {http://dx.doi.org/10.1016/0370-1573(87)90144-X} {Stochastic
  quantization}.
\newblock {\em Physics Reports} {\bf 152}, 227 (1987).

\bibitem{chaturvedi_ward_1984}
S.~Chaturvedi, A.~K. Kapoor, and V.~Srinivasan.
\newblock \href {http://dx.doi.org/10.1007/BF01318417} {Ward {Takahashi}
  identities and fluctuation-dissipation theorem in a superspace formulation of
  the {Langevin} equation}.
\newblock {\em Z. Physik B Condensed Matter} {\bf 57}, 249 (1984).

\bibitem{gozzi_onsager_1984}
E.~Gozzi.
\newblock \href {http://dx.doi.org/10.1103/PhysRevD.30.1218} {Onsager principle
  of microscopic reversibility and supersymmetry}.
\newblock {\em Phys. Rev. D} {\bf 30}, 1218 (1984).

\bibitem{zimmer_fluctuations_1993}
M.~F. Zimmer.
\newblock \href {http://dx.doi.org/10.1007/BF01054348} {Fluctuations in
  nonequilibrium systems and broken supersymmetry}.
\newblock {\em J. Stat. Phys.} {\bf 73}, 751 (1993).

\bibitem{gawedzki_critical_1986}
K.~Gawedzki and A.~Kupiainen.
\newblock \href {http://dx.doi.org/10.1016/0550-3213(86)90364-0} {Critical
  behaviour in a model of stationary flow and supersymmetry breaking}.
\newblock {\em Nuclear Physics B} {\bf 269}, 45 (1986).

\bibitem{trimper_supersymmetry_1990}
S.~Trimper.
\newblock \href {http://dx.doi.org/10.1088/0305-4470/23/4/008} {Supersymmetry
  breaking for dynamical systems}.
\newblock {\em J. Phys. A: Math. Gen.} {\bf 23}, L169 (1990).

\bibitem{agarwal_fluctuation-dissipation_1972}
G.~S. Agarwal.
\newblock \href {http://dx.doi.org/10.1007/BF01391621} {Fluctuation-dissipation
  theorems for systems in non-thermal equilibrium and applications}.
\newblock {\em Z Physik A Hadrons and nuclei} {\bf 252}, 25 (1972).

\bibitem{speck_restoring_2006}
T.~Speck and U.~Seifert.
\newblock \href {http://dx.doi.org/10.1209/epl/i2005-10549-4} {Restoring a
  fluctuation-dissipation theorem in a nonequilibrium steady state}.
\newblock {\em EPL (Europhysics Letters)} {\bf 74}, 391 (2006).

\bibitem{prost_generalized_2009}
J.~Prost, J.-F. Joanny, and J.~M.~R. Parrondo.
\newblock \href {http://dx.doi.org/10.1103/PhysRevLett.103.090601} {Generalized
  fluctuation-dissipation theorem for steady-state systems}.
\newblock {\em Phys. Rev. Lett.} {\bf 103}, 090601 (2009).

\bibitem{Janssen1976}
H.-K. Janssen.
\newblock \href {http://dx.doi.org/10.1007/BF01316547} {{On a Lagrangean for
  classical field dynamics and renormalization group calculations of dynamical
  critical properties}}.
\newblock {\em Z. Physik B Condensed Matter} {\bf 23}, 377 (1976).

\bibitem{janssen_field-theoretic_1979}
H.-K. Janssen.
\newblock \href {http://dx.doi.org/10.1007/3-540-09523-3_2} {Field-theoretic
  method applied to critical dynamics}.
\newblock In Charles~P. Enz, editor, {\em Dynamical {Critical} {Phenomena} and
  {Related} {Topics}} Lecture {Notes} in {Physics} pages~25 Berlin, Heidelberg
  1979. Springer.

\bibitem{dominicis_techniques_1976}
C.~De Dominicis.
\newblock \href {http://dx.doi.org/10.1051/jphyscol:1976138} {Techniques de
  renormalisation de la th{\'e}orie des champs et dynamique des
  ph{\'e}nom{\`e}nes critiques}.
\newblock {\em J. Phys. Colloques} {\bf 37}, 247 (1976).

\bibitem{DeDominicis1978}
C.~De~Dominicis and L.~Peliti.
\newblock \href {http://link.aps.org/doi/10.1103/PhysRevB.18.353}
  {{Field-theory renormalization and critical dynamics above ${T}_{c}$: Helium,
  antiferromagnets, and liquid-gas systems}}.
\newblock {\em Phys. Rev. B} {\bf 18}, 353 (1978).

\bibitem{Martin1973}
P.~C. Martin, E.~D. Siggia, and H.~A. Rose.
\newblock \href {http://link.aps.org/doi/10.1103/PhysRevA.8.423} {Statistical
  dynamics of classical systems}.
\newblock {\em Phys. Rev. A} {\bf 8}, 423 (1973).

\bibitem{onsager_fluctuations_1953}
L.~Onsager and S.~Machlup.
\newblock \href {http://dx.doi.org/10.1103/PhysRev.91.1505} {Fluctuations and
  irreversible processes}.
\newblock {\em Phys. Rev.} {\bf 91}, 1505 (1953).

\bibitem{machlup_fluctuations_1953II}
S.~Machlup and L.~Onsager.
\newblock \href {http://dx.doi.org/10.1103/PhysRev.91.1512} {Fluctuations and
  irreversible process. {II}. {Systems} with kinetic energy}.
\newblock {\em Phys. Rev.} {\bf 91}, 1512 (1953).

\bibitem{kardar_dynamic_1986}
M.~Kardar, G.~Parisi, and Y.-C. Zhang.
\newblock \href {http://dx.doi.org/10.1103/PhysRevLett.56.889} {Dynamic scaling
  of growing interfaces}.
\newblock {\em Phys. Rev. Lett.} {\bf 56}, 889 (1986).

\bibitem{risken_fokker-planck_1996}
H.~Risken.
\newblock {\em The {Fokker}-{Planck} equation: methods of solution and
  applications}.
\newblock Number v. 18 in Springer series in synergetics. Springer-Verlag New
  York 2nd ed edition 1996.

\bibitem{graham_covariant_1977}
R.~Graham.
\newblock \href {http://dx.doi.org/10.1007/BF01570750} {Covariant formulation
  of non-equilibrium statistical thermodynamics}.
\newblock {\em Z. Physik B Condensed Matter} {\bf 26}, 397 (1977).

\bibitem{eyink_hydrodynamics_1996}
G.~L. Eyink, J.~L. Lebowitz, and H.~Spohn.
\newblock \href {http://dx.doi.org/10.1007/BF02183738} {Hydrodynamics and
  fluctuations outside of local equilibrium: driven diffusive systems}.
\newblock {\em J. Stat. Phys.} {\bf 83}, 385 (1996).

\bibitem{Langouche81}
F.~Langouche, D.~Roekaerts, and E.~Tirapegui.
\newblock General Langevin equations and functional integration.
\newblock In {\em Field Theory, Quantization and Statistical Physics: in Memory
  of Bernard Jouvet}. Springer Dordrecht 1981.

\bibitem{itami_universal_2017}
M.~Itami and S-i. Sasa.
\newblock \href {http://dx.doi.org/10.1007/s10955-017-1738-6} {Universal form
  of stochastic evolution for slow variables in equilibrium systems}.
\newblock {\em J. Stat. Phys.} {\bf 167}, 46 (2017).

\bibitem{Cugliandolo-Lecomte17a}
L.~F. Cugliandolo and V.~Lecomte.
\newblock \href {http://dx.doi.org/10.1088/1751-8121/aa7dd6} {Rules of calculus
  in the path integral representation of white noise {Langevin} equations: the
  {Onsager}--{Machlup} approach}.
\newblock {\em J. Phys. A: Math. Theor.} {\bf 50}, 345001 (2017).

\bibitem{cugliandolo_building_2019}
L.~F. Cugliandolo, V.~Lecomte, and F.~van Wijland.
\newblock \href {http://dx.doi.org/10.1088/1751-8121/ab3ad5} {Building a
  path-integral calculus: a covariant discretization approach}.
\newblock {\em J. Phys. A: Math. Theor.} {\bf 52}, 50LT01 (2019).

\bibitem{arenas_supersymmetric_2012}
Z.~G. Arenas and D.~G. Barci.
\newblock \href {http://dx.doi.org/10.1103/PhysRevE.85.041122} {Supersymmetric
  formulation of multiplicative white-noise stochastic processes}.
\newblock {\em Phys. Rev. E} {\bf 85}, 041122 (2012).

\bibitem{arenas_hidden_2012}
Z.~G. Arenas and D.~G. Barci.
\newblock \href {http://dx.doi.org/10.1088/1742-5468/2012/12/P12005} {Hidden
  symmetries and equilibrium properties of multiplicative white-noise
  stochastic processes}.
\newblock {\em J. Stat. Mech.} {\bf 2012}, P12005 (2012).

\bibitem{berezin2012method}
F.~A. Berezin.
\newblock {\em The method of second quantization}.
\newblock Academic Press 1966.

\bibitem{Note1}
It is more rigorously defined as $\varphi _i[h+h^1]=\varphi _i[h]+\varphi
  _i[h]\protect \text {\protect \textbf {'}}h^1 + o(h^1)$, implying
  that~$\partial _t(\varphi [h])=\varphi \protect \text {\protect \textbf
  {'}}\partial _t h$, and $\varphi _i[h+\varepsilon \Psi ]=\varphi
  _i[h]+\varepsilon \varphi _i[h]\protect \text {\protect \textbf {'}}\Psi $.
  One has for instance for $\eta _i[h]$ defined in Eq.~(\ref {eq:PhOMetaih}):
  $\eta _i\sns \pp [h]\Psi =\big (\delta _{ij}\partial _t-\partial _jf_i[h]\big
  )\Psi _{\sns j}$.

\bibitem{nicolai_supersymmetry_1980}
H.~Nicolai.
\newblock \href {http://dx.doi.org/10.1016/0550-3213(80)90460-5} {Supersymmetry
  and functional integration measures}.
\newblock {\em Nuclear Physics B} {\bf 176}, 419 (1980).

\bibitem{nicolai_new_1980}
H.~Nicolai.
\newblock \href {http://dx.doi.org/10.1016/0370-2693(80)90138-0} {On a new
  characterization of scalar supersymmetric theories}.
\newblock {\em Physics Letters B} {\bf 89}, 341 (1980).

\bibitem{nicolai_functional_1982}
H.~Nicolai.
\newblock \href {http://dx.doi.org/10.1016/0370-2693(82)90570-6} {On the
  functional integration measure of supersymmetric {Yang}-{Mills} theories}.
\newblock {\em Physics Letters B} {\bf 117}, 408 (1982).

\bibitem{cecotti_stochastic_1983}
S.~Cecotti and L.~Girardello.
\newblock \href {http://dx.doi.org/10.1016/0003-4916(83)90172-0} {Stochastic
  and parastochastic aspects of supersymmetric functional measures: {A} new
  non-perturbative approach to supersymmetry}.
\newblock {\em Annals of Physics} {\bf 145}, 81 (1983).

\bibitem{Note2}
Hence, explicitly: $ \Psib _{\sns i} \tilde \eta _i\sns \pp ^\dag \Psi = \Psib
  _{\sns i}(\delta _{ij}\partial _t +\partial _{ij}\mathcal H+\partial _i
  g_j)\Psi _{\sns j} $.

\bibitem{graham_statistical_1973}
R.~Graham.
\newblock \href {http://dx.doi.org/10.1007/978-3-662-40468-3_1} {Statistical
  theory of instabilities in stationary nonequilibrium systems with
  applications to lasers and nonlinear optics}.
\newblock In G.~H{\"o}hler, editor, {\em Springer {Tracts} in {Modern}
  {Physics}: {Ergebnisse} der exakten {Naturwissenschaftenc}; {Volume} 66}
  pages~1. Springer Berlin, Heidelberg 1973.

\bibitem{LaRoTi79}
F.~Langouche, D.~Roekaerts, and E.~Tirapegui.
\newblock \href {http://dx.doi.org/10.1007/BF02739307} {Functional integrals
  and the {Fokker}-{Planck} equation}.
\newblock {\em Il Nuovo Cimento B} {\bf 53}, 135 (1979).

\bibitem{Tirapegui82}
F.~Langouche, D.~Roekaerts, and E.~Tirapegui.
\newblock {\em Functional integration and semiclassical expansions}.
\newblock Kluwer Academic Publishers Dordrecht 1982.

\bibitem{janssen_renormalized_1992}
H.~K. Janssen.
\newblock \href
  {http://www.worldscientific.com/doi/abs/10.1142/9789814355872_0007} {On the
  renormalized field theory of nonlinear critical relaxation}.
\newblock In {\em From {Phase} {Transitions} to {Chaos}} pages~68. World
  Scientific 1992.

\bibitem{lau_state-dependent_2007}
A.~W.~C. Lau and T.~C. Lubensky.
\newblock \href {http://dx.doi.org/10.1103/PhysRevE.76.011123} {State-dependent
  diffusion: thermodynamic consistency and its path integral formulation}.
\newblock {\em Phys. Rev. E} {\bf 76}, 011123 (2007).

\bibitem{arenas_functional_2010}
Z.~G. Arenas and D.~G. Barci.
\newblock \href {http://dx.doi.org/10.1103/PhysRevE.81.051113} {Functional
  integral approach for multiplicative stochastic processes}.
\newblock {\em Phys. Rev. E} {\bf 81}, 051113 (2010).

\bibitem{aron_dynamical_2016}
C.~Aron, D.~G. Barci, L.~F. Cugliandolo, Z.~G. Arenas, and G.~S. Lozano.
\newblock \href {http://dx.doi.org/10.1088/1742-5468/2016/05/053207} {Dynamical
  symmetries of {Markov} processes with multiplicative white noise}.
\newblock {\em J. Stat. Mech.} {\bf 2016}, 053207 (2016).

\bibitem{Note3}
Indeed, discretizing with a time step $\Delta t$, one has $ \eta _i[h]_t =
  \frac {h_{i,t+\Delta t}-h_{i,t}}{\Delta t} - f_i[h] \big |_{h=\frac
  12(h_{i,t+\Delta t}+h_{i,t})} $ where the time is in index (and
  discretization is Stratonovich). Hence the matrix of coordinates $(i,t;j,t')$
  in the definition of the Jacobian after Eq.~(\ref {eq:PhOMetaih}) is upper
  triangular in the time direction (this is causality), so that only its
  equal-time components matter. Importantly, since the time-discrete Langevin
  equation is read as $h_{t+\Delta t}$ function of $h_t$ and $\eta _t$, one
  must pay attention that the change of variables is between $h_{t+\Delta t}$
  and $\eta _t$. Its Jacobian is thus $\partial \eta _i[h]_t/\partial
  h_{j,t+\Delta t} = \protect \frac {1}{\Delta t}\delta _{ij} - \protect \frac
  12 \partial _j f_i[h] $. Factorizing by $\protect \frac {1}{\Delta t}$ [which
  yields a field-independent normalization factor of the Jacobian], using the
  formula $\protect \qopname \relax o{log}\protect \qopname \relax m{det}=
  \protect \operatorname {tr}\protect \qopname \relax o{log}$, one thus obtains
  $ \log \big |\!\frac {\delta \eta }{\delta h}\!\big |= \sum _t \tr \log
  (\mathbf {1}-\frac 12 \Delta t f_i\pp [h]) $. Expanding at small $\Delta t$,
  one recovers Eq.\protect \:(\ref {eq:JacExpl}).

\bibitem{Note4}
Denoting by $X^{\protect \text {\relax \protect \fontsize {5}{6}\protect
  \selectfont {s}}}_t=\protect \frac {X_{t+\Delta t}+X_t}{2}$ the Stratonovich
  discretization, $ \int _t \Psib _{\sns i} \eta _i\sns \pp [h]\Psi = \int _t
  \Psib _{\sns i} (\delta _{ij}\partial _t-\p _j\sns f_i[h])\Psi _{\sns j} $
  must be discretized as $ \sum _t \!\Delta t\sms \Psib _{\sns i,t+\Delta t}
  \big (\frac {\Psi _{\sns i,t+\Delta t}-\Psi _{\sns i,t}}{\Delta t}-\p _j\sns
  f_i[h^\St _t]\Psi ^\St _{\sns j,t}\big ) = \sum _{tt'}\! \Psib _{\sns i,t'}
  M_{i,t';j,t}\Psi _{\sns j,t} $ with the matrix elements given by $
  M_{i,t'\!;j,t} = \delta _{ij} (\delta _{t'\!,t}-\delta _{t'\!,t+\Delta t}) -
  \!\Delta t\sms \p _j\sns f_i[h_{t'}] \frac {\delta _{t'\!,t}+\delta
  _{t'\!,t+\Delta t}}{2} $. As the Grassmann integral yields the determinant of
  $M$, and as $M$ is triangular in the time coordinate, only the diagonal
  $t'=t$ matters and $ \det M = \sum _t \det (\delta _{ij}-\!\Delta t\sms
  \partial _jf_i[h]) $. One thus recovers the Jacobian~\cite {Note3}.

\bibitem{Note5}
This explains why one cannot transform $S_{\protect \text {\relax \protect
  \fontsize {5}{6}\protect \selectfont {SUSY}}}$ into $\protect \smash
  {S_{\protect \text {\relax \protect \fontsize {5}{6}\protect \selectfont
  {SUSY}}}^{\protect \smash \protect \dag }}$: these actions contain the same
  information after integrating on the Grassmann fields, but not before.

\bibitem{inpreparation}
B.~Marguet, E.~Agoritsas, L.~Canet, and V.~Lecomte.
\newblock (In preparation) 2021.

\bibitem{Note6}
With the notations of~\cite {Note4}, we have that $ \langle \Psib _{i,t'} \Psi
  _{j,t}\rangle ^{\!\dag } = {(M^T)^{-1}}_{i,t';j,t} $, but $M$ is lower
  triangular in the time direction, so that $(M^T)^{-1}$ is upper triangular.
  This implies the causality $ \langle \Psib _{i,t'} \Psi _{j,t}\rangle
  ^{\!\dag } =0 $ for $t'>t$.

\bibitem{falcioni_correlation_1990}
Massimo Falcioni, Stefano Isola, and Angelo Vulpiani.
\newblock \href {http://dx.doi.org/10.1016/0375-9601(90)90137-D} {Correlation
  functions and relaxation properties in chaotic dynamics and statistical
  mechanics}.
\newblock {\em Physics Letters A} {\bf 144}, 341 (1990).

\bibitem{crooks_path-ensemble_2000}
G.~E. Crooks.
\newblock \href {http://dx.doi.org/10.1103/PhysRevE.61.2361} {Path-ensemble
  averages in systems driven far from equilibrium}.
\newblock {\em Phys. Rev. E} {\bf 61}, 2361 (2000).

\bibitem{chetrite_fluctuation_2008}
R.~Chetrite, G.~Falkovich, and K.~Gawedzki.
\newblock \href {http://dx.doi.org/10.1088/1742-5468/2008/08/P08005}
  {Fluctuation relations in simple examples of non-equilibrium steady states}.
\newblock {\em J. Stat. Mech.} {\bf 2008}, P08005 (2008).

\bibitem{baiesi_fluctuations_2009}
M.~Baiesi, C.~Maes, and B.~Wynants.
\newblock \href {http://dx.doi.org/10.1103/PhysRevLett.103.010602}
  {Fluctuations and response of nonequilibrium states}.
\newblock {\em Phys. Rev. Lett.} {\bf 103}, 010602 (2009).

\bibitem{seifert_fluctuation-dissipation_2010}
U.~Seifert and T.~Speck.
\newblock \href {http://dx.doi.org/10.1209/0295-5075/89/10007}
  {Fluctuation-dissipation theorem in nonequilibrium steady states}.
\newblock {\em EPL} {\bf 89}, 10007 (2010).

\bibitem{verley_modified_2011}
G.~Verley, K.~Mallick, and D.~Lacoste.
\newblock \href {http://dx.doi.org/10.1209/0295-5075/93/10002} {Modified
  fluctuation-dissipation theorem for non-equilibrium steady states and
  applications to molecular motors}.
\newblock {\em EPL} {\bf 93}, 10002 (2011).

\bibitem{dal_cengio_fluctuationdissipation_2021}
Sara Dal~Cengio, Demian Levis, and Ignacio Pagonabarraga.
\newblock \href {http://dx.doi.org/10.1088/1742-5468/abee22}
  {Fluctuation–dissipation relations in the absence of detailed balance:
  formalism and applications to active matter}.
\newblock {\em Journal of Statistical Mechanics: Theory and Experiment} {\bf
  2021}, 043201 (2021).

\bibitem{canet2011nonperturbative}
L.~Canet, H.~Chat{\'e}, B.~Delamotte, and N.~Wschebor.
\newblock \href {http://dx.doi.org/10.1103/PhysRevE.84.061128} {Nonperturbative
  renormalization group for the {Kardar}-{Parisi}-{Zhang} equation: general
  framework and first applications}.
\newblock {\em Phys. Rev. E} {\bf 84}, 061128 (2011).

\bibitem{chernyak_path-integral_2006}
V.~Y. Chernyak, M.~Chertkov, and C.~Jarzynski.
\newblock \href {http://dx.doi.org/10.1088/1742-5468/2006/08/P08001}
  {Path-integral analysis of fluctuation theorems for general {Langevin}
  processes}.
\newblock {\em J. Stat. Mech.} {\bf 2006}, P08001 (2006).

\bibitem{brian_anderson_reverse-time_1982}
Brian Anderson.
\newblock \href {http://dx.doi.org/10.1016/0304-4149(82)90051-5} {Reverse-time
  diffusion equation models}.
\newblock {\em Stochastic Processes and their Applications} {\bf 12}, 313
  (1982).

\bibitem{Note7}
The Onsager--Machlup actions take a simple form: $ S_\OM =\int _t\! \big \{
  \frac {\eta D^{\sns -\sns 1\sns }\eta }{4T}-\Psib \eta \pp \Psi \big \} $
  and\protect \, $ \mathbb S^\dag _\OM =\int _t\! \big \{ \frac {\bar \eta
  D^{\sns -\sns 1\sns }\bar \eta }{4T}+\Psi \bar \eta \pp \Psib \big \} $, with
  $S_{\protect \text {\relax \protect \fontsize {5}{6}\protect \selectfont
  {OM}}}$ verifying the BRST SUSY~(\ref {eq:BRS_transformation}), and $\protect
  \mathbb S^\protect \dag _{\protect \text {\relax \protect \fontsize
  {5}{6}\protect \selectfont {OM}}}$ being invariant by the PS SUSY $\delta
  h=\varepsilon T\mkern 1.6mu\protect \overline {\mkern -1.6mu\Psi \mkern
  -1.6mu}\mkern 1.6mu$, $\delta \Psi =-\protect \frac \varepsilon 2 D^{\kern
  -.75pt-\kern -.75pt1\kern -.75pt}\protect \bar \eta $, $\delta \mkern
  1.6mu\protect \overline {\mkern -1.6mu\Psi \mkern -1.6mu}\mkern 1.6mu=0$
  corresponding to PS$_{1,2}$.

\bibitem{mallick_field-theoretic_2011}
K.~Mallick, M.~Moshe, and H.~Orland.
\newblock \href {http://dx.doi.org/10.1088/1751-8113/44/9/095002} {A
  field-theoretic approach to non-equilibrium work identities}.
\newblock {\em J. Phys. A: Math. Theor.} {\bf 44}, 095002 (2011).

\bibitem{spitzer_frank_interaction_1970}
F.~Spitzer.
\newblock \href {http://dx.doi.org/10.1016/0001-8708(70)90034-4} {Interaction
  of {Markov} processes}.
\newblock {\em Advances in Mathematics} {\bf 5}, 246 (1970).

\bibitem{levine_zero-range_2005}
E.~Levine, D.~Mukamel, and G.~M. Sch{\"u}tz.
\newblock \href {http://dx.doi.org/10.1007/s10955-005-7000-7} {Zero-range
  process with open boundaries}.
\newblock {\em J. Stat. Phys.} {\bf 120}, 759 (2005).

\bibitem{evans_nonequilibrium_2005}
M.~R. Evans and T.~Hanney.
\newblock \href {http://dx.doi.org/10.1088/0305-4470/38/19/R01} {Nonequilibrium
  statistical mechanics of the zero-range process and related models}.
\newblock {\em J. Phys. A: Math. Gen.} {\bf 38}, R195 (2005).

\bibitem{guioth_mass_2017}
J.~Guioth and E.~Bertin.
\newblock \href {http://dx.doi.org/10.1088/1742-5468/aa6de2} {A mass transport
  model with a simple non-factorized steady-state distribution}.
\newblock {\em J. Stat. Mech.} {\bf 2017}, 063201 (2017).

\bibitem{schutz_1_2001}
G.~M. Sch{\"u}tz.
\newblock \href
  {http://www.sciencedirect.com/science/article/pii/S106279010180015X}
  {{Exactly} {Solvable} {Models} for {Many}-{Body} {Systems} {Far} from
  {Equilibrium}}.
\newblock In C.~Domb {and} J.~L. Lebowitz, editor, {\em Phase {Transitions} and
  {Critical} {Phenomena}} volume~19 pages~1. Academic Press 2001.

\bibitem{giardina_duality_2009}
Cristian Giardinà, Jorge Kurchan, Frank Redig, and Kiamars Vafayi.
\newblock \href {http://dx.doi.org/10.1007/s10955-009-9716-2} {Duality and
  {Hidden} {Symmetries} in {Interacting} {Particle} {Systems}}.
\newblock {\em Journal of Statistical Physics} {\bf 135}, 25 (2009).

\bibitem{frassek_non-compact_2020}
Rouven Frassek, Cristian Giardinà, and Jorge Kurchan.
\newblock \href {http://dx.doi.org/10.1007/s10955-019-02375-4} {Non-compact
  {Quantum} {Spin} {Chains} as {Integrable} {Stochastic} {Particle}
  {Processes}}.
\newblock {\em Journal of Statistical Physics} {\bf 180}, 135 (2020).

\bibitem{bertini_macroscopic_2015}
L.~Bertini, A.~De~Sole, D.~Gabrielli, G.~Jona-Lasinio, and C.~Landim.
\newblock \href {http://dx.doi.org/10.1103/RevModPhys.87.593} {Macroscopic
  fluctuation theory}.
\newblock {\em Rev. Mod. Phys.} {\bf 87}, 593 (2015).

\bibitem{freidlin_random_2012}
Mark~I. Freidlin and Alexander~D. Wentzell.
\newblock \href {//www.springer.com/gp/book/9783642258466} {{\em Random
  {Perturbations} of {Dynamical} {Systems}}}.
\newblock Grundlehren der mathematischen {Wissenschaften}. Springer-Verlag
  Berlin Heidelberg 3 edition 2012.

\bibitem{graham_weak-noise_1984}
R.~Graham and T.~Tél.
\newblock \href {http://dx.doi.org/10.1007/BF01010830} {On the weak-noise limit
  of {Fokker}-{Planck} models}.
\newblock {\em Journal of Statistical Physics} {\bf 35}, 729 (1984).

\bibitem{graham_weak-noise_1985}
R.~Graham and T.~Tél.
\newblock \href {http://dx.doi.org/10.1103/PhysRevA.31.1109} {Weak-noise limit
  of {Fokker}-{Planck} models and nondifferentiable potentials for dissipative
  dynamical systems}.
\newblock {\em Physical Review A} {\bf 31}, 1109 (1985).

\bibitem{kamenev_field_2011}
Alex Kamenev.
\newblock {\em Field theory of non-equilibrium systems}.
\newblock Cambridge University Press Cambridge ; New York 2011.

\bibitem{drummond_stochastic_2012}
I.~T. Drummond and R.~R. Horgan.
\newblock \href {http://dx.doi.org/10.1088/1751-8113/45/9/095005} {Stochastic
  processes, slaves and supersymmetry}.
\newblock {\em J. Phys. A: Math. Theor.} {\bf 45}, 095005 (2012).

\bibitem{dean_equilibrium_2005}
D.~S. Dean, I.~T. Drummond, R.~R. Horgan, and S.~N. Majumdar.
\newblock \href {http://dx.doi.org/10.1103/PhysRevE.71.031103} {Equilibrium
  statistics of a slave estimator in {Langevin} processes}.
\newblock {\em Phys. Rev. E} {\bf 71}, 031103 (2005).

\bibitem{ovchinnikov_introduction_2016}
I.~V. Ovchinnikov.
\newblock \href {http://dx.doi.org/10.3390/e18040108} {Introduction to
  supersymmetric theory of stochastics}.
\newblock {\em Entropy} {\bf 18}, 108 (2016).

\bibitem{wegner_supermathematics_2016}
F.~Wegner.
\newblock \href {http://dx.doi.org/10.1007/978-3-662-49170-6} {{\em
  Supermathematics and its applications in statistical physics}} volume 920 of
  {\em Lecture {Notes} in {Physics}}.
\newblock Springer Berlin Heidelberg 2016.

\end{thebibliography}

\end{document}